\documentclass[sigconf, screen]{acmart}
\usepackage[]{hyperref}
\usepackage{subcaption,amsmath,amsthm,amsfonts,pifont,bm,algorithm,xcolor,enumitem,cleveref,graphicx}
\usepackage{xcolor}

\graphicspath{{images/}{../images/}}
\crefname{section}{§}{§§}

\copyrightyear{2025}
\acmYear{2025}
\setcopyright{cc}
\setcctype{by}
\acmConference[SC '25]{The International Conference for High Performance Computing, Networking, Storage and Analysis}{November 16--21, 2025}{St Louis, MO, USA}
\acmBooktitle{The International Conference for High Performance Computing, Networking, Storage and Analysis (SC '25), November 16--21, 2025, St Louis, MO, USA}
\acmDOI{10.1145/3712285.3759784}
\acmISBN{979-8-4007-1466-5/2025/11}

\begin{document}

\title{Hetis: Serving LLMs in Heterogeneous GPU Clusters with Fine-grained and Dynamic Parallelism}

\author{Zizhao Mo}
\affiliation{
	\institution{University of Macau}
	\country{Macau SAR, China}
}
\email{yc17461@connect.um.edu.mo}

\author{Jianxiong Liao}
\affiliation{
	\institution{Sun Yat-sen University}
	\country{China}
}
\email{liaojx9@mail2.sysu.edu.cn}

\author{Huanle Xu}
\affiliation{
	\institution{University of Macau}
	\country{Macau SAR, China}
}
\email{huanlexu@um.edu.mo}
\authornote{Corresponding author}

\author{Zhi Zhou}
\affiliation{
	\institution{Sun Yat-sen University}
	\country{China}
}
\email{zhouzhi9@mail.sysu.edu.cn}

\author{Chengzhong Xu}
\affiliation{
	\institution{University of Macau}
	\country{Macau SAR, China}
}
\email{czxu@um.edu.mo}

\begin{abstract}
The significant resource demands in LLM serving prompts production clusters to fully utilize heterogeneous hardware by partitioning LLM models across a mix of high-end and low-end GPUs. However, existing parallelization approaches often struggle to scale efficiently in heterogeneous environments due to their coarse-grained and static parallelization strategies.

In this paper, we introduce Hetis, a new LLM system tailored for heterogeneous GPU clusters. Hetis addresses two critical challenges: (1) memory inefficiency caused by the mismatch between memory capacity and computational power in heterogeneous devices, and (2) computational inefficiency arising from performance gaps across different LLM modules. To tackle these issues, Hetis employs a fine-grained and dynamic parallelism design. Specifically, it selectively parallelizes compute-intensive operations to reduce latency and dynamically distributes Attention computations to low-end GPUs at a head granularity, leveraging the distinct characteristics of each module. Additionally, Hetis features an online load dispatching policy that continuously optimizes serving performance by carefully balancing network latency, computational load, and memory intensity.  Evaluation results demonstrate that Hetis can improve serving throughput by up to $2.25\times$ and reduce latency by $1.49\times$ compared to existing systems.
\end{abstract}

\maketitle

\section{Introduction}
Nowadays, large language models (LLMs) have widely proven their superiority across various domains, such as question answering, translation, and code generation \cite{llama, llama2, gpt4, t5, falcon, opt}. Inspired by the scaling law, which states that the performance of deep learning models generally improves with the number of parameters \cite{fewshot, scalinglaw}, it has become common practice for companies to race towards deploying LLMs with increasingly massive parameter scales.

Serving giant LLM models is typically associated with high resource requirements. On one hand, the inference process involves a large amount of floating-point calculations within modules such as MLP (Multi-Layer Perceptron)~\cite{attention}, heavily relying on the parallel computation capabilities provided by the SM cores in GPUs. On the other hand, hosting the model parameters and intermediate activation tensors in GPUs necessitates a substantial amount of memory space. Furthermore, generative LLMs are characterized by their autoregressive token generation process, where the output of the next token depends on the intermediate results of all previous tokens during Attention computation~\cite{attention, orca}.  Consequently, state-of-the-art serving systems leverage the KV (key-value) cache technique to store the key and value vectors, thereby avoiding redundant computation at the cost of increased memory consumption~\cite{orca, pagedattention, alpaserve, dejavu, spotserve}. For example, decoding a single sequence alone with a length of 10k in a LLaMA2-13B model requires more than 8 GB of memory~\cite{infinitellm}.


The high resource demands of LLM serving necessitate fully harnessing the available resources within a cluster. Modern production clusters, comprising a diverse array of devices due to the rapid evolution of hardware, inherently feature heterogeneous infrastructures~\cite{gavel, allox, gandivafair, sia, heet, oef, fft}. Therefore, effectively harnessing both high- and low-end GPUs is crucial for enhancing LLM serving performance. However, the processing speed of various modules within an LLM model varies significantly across heterogeneous GPUs, posing challenges in efficiently utilizing these resources. Notably, processing the MLP module on a P100 GPU results in a substantial increase in processing time, extending it by a factor of 24.5 compared to an A100 GPU for the same requests. This performance disparity can easily lead to resource underutilization, with high-end GPUs left idle while waiting for hidden states to be gathered from low-end ones.  Furthermore, memory heterogeneity across different GPU types presents an additional challenge, as the limited memory capacity of low-end GPUs necessitates advanced model parallelization to accommodate KV caches.

State-of-the-art heterogeneity-aware LLM serving systems~\cite{llmpq, melange, hexgen,splitwise} employ uneven parallelization of computations across GPUs, taking into account the heterogeneous capabilities of different devices. These systems strategically  harmonize inference time and prevent the overall performance from being adversely affected by the slowest GPU, via leveraging a combination of data-, pipeline-, and tensor-parallelism. However, we have identified fundamental limitations in these strategies, stemming from the inability to differentiate between various modules of LLM serving and the reliance on static parallelization approaches. First, employing existing parallelization schemes leads to low memory efficiency. For instance, the Prefill-Decode disaggregation scheme like Splitwise~\cite{splitwise} consumes substantial memory by hosting extra copies of model parameters. In contrast, asymmetric parallelization optimized the computational power gaps across heterogeneous devices, like Hexgen~\cite{hexgen}, results in significant unused memory space under fluctuating request arrivals. Second, current heterogeneity-aware systems inadvertently introduce computation inefficiencies. The phase-wise disaggregation approach relies on high-end devices solely for the computation-intensive prefill phase, resulting in prolonged decoding, especially with long contexts. Moreover, the significant difference in computation intensity across MLP and Attention modules can easily lead to a mismatch in computation time under the unified and static parallelization scheme within Hexgen. Worse still, the benefit of latency reduction under this significant performance gap may be readily offset by the collective communication overhead.

In this paper, we present Hetis, a new LLM serving system designed to minimize inference latency while maintaining high throughput in heterogeneous clusters.
Motivated by the opportunities to selectively distribute the computationally intensive operations across heterogeneous GPUs and parallelize the Attention computation for fine-grained and dynamic parallelism, Hetis introduces a novel parallelization scheme.  At its core, Hetis features a primary worker parallelism mechanism that optimizes the computation cost of dense modules on a carefully selected subset of GPU devices, identified through a rigorous optimization problem formulation. The remaining devices are reserved exclusively for Attention computation. Furthermore, Hetis incorporates a dynamic Attention parallelization mechanism, which facilitates flexible parallelization of Attention computation for inference requests along the head dimension. This mechanism adapts seamlessly to highly fluctuating request arrivals, minimizing communication overhead during the distributed decoding process.

On top of the parallelization scheme, Hetis develops an online request-level dispatching algorithm to optimize resource utilization across the entire cluster, thereby minimizing inference latency and enhancing throughput. Specifically, this algorithm  dynamically allocates Attention heads for incoming requests across heterogeneous GPUs in real-time, taking into account factors including network delay, memory space, and computation capabilities across multiple workers. Moreover, Hetis designs a meticulous load balancing scheme
tailored for the generation of long contexts, effectively mitigating uneven memory usage and redundant cache transmission. To facilitate these new algorithm designs, Hetis incorporates head-specific KV cache management and computation, and interference-free cache migration protocols.

We have developed a prototype of Hetis by building upon the popular vLLM framework~\cite{pagedattention} and performed comprehensive experiments in our heterogeneous cluster to assess its performance. Experiment results showcase Hetis can reduce inference latency by up to $1.49\times$. Furthermore, Hetis proves a remarkable improvement in throughput, sustaining up to $2.25\times$ higher request rate than state-of-the-art heterogeneity-aware systems. In summary, this paper presents the following contributions:
\begin{itemize}
    \item We have conducted an in-depth analysis to investigate the impact of GPU heterogeneity on state-of-the-art LLM serving systems. Our analysis highlights the significance of implementing a more fine-grained and dynamic parallelization solution  to fully leverage the heterogeneous GPU resources.
    \item We have introduced a novel parallelism design dedicated for LLM serving over heterogeneous GPU clusters. By this parallelism, we can achieve a more efficient and fine-grained distribution of computation and memory usage at the module level, thereby maximizing resource utilization.
    \item We have proposed a scalable load dispatching policy to seamlessly adapt to dynamic request arrivals and long context generation to maximize the benefits of our new parallelism mechanisms. This is facilitated by efficient optimization and explicit quantification of network, computation, and memory conditions.
\end{itemize}

\section{Background and Motivation}
\subsection{LLM Serving Characteristics}
Today's LLM models are typically built upon multiple Transformer layers~\cite{attention}, in which each layer consists of the identical structure but different model parameters. To be specific, each layer translates the hidden states from the output of the previous layer or input embeddings to \textit{Q}, \textit{K}, \textit{V} vectors in the \textit{QKV} module, and then calculates the Attention value to focus on diverse parts of input sequences in the \textit{Attention} module, which is crucial for understanding context and relationships within the data. To enhance the expressiveness, Attention computation is manipulated in the \textit{head} granularity, i.e., a partial chunk of hidden state. Finally, the attention scores are projected and then delivered to the dense  module  \textit{MLP}, which further captures complex patterns from the attention output.

The LLM inference process can be divided into two phases, based on their computational characteristics. The first phase, known as the \textit{Prefill Phase}, is responsible for generating the initial output from the entire prompt context at one shot, while the subsequent \textit{Decode Phase} autoregressively produces one token at a time from the prompt and newly generated tokens until terminated. As the \textit{K} and \textit{V} matrixes of historic tokens required by generating each token in the decode phase consume substantial amount of GPU memory, the memory wall in GPU, i.e., the limited GPU memory capacity, turns out to be the most critical concern in the LLM inference, drawing forth a body of works that pursue high memory efficiency~\cite{pagedattention}.

LLM serving is also well-known for its dynamic nature. Similar to inference systems for other deep learning tasks~\cite{shepherd, clockwork, mark, lazybatching, batch, nexus}, the arrival rate of serving requests for LLM service experiences significant fluctuations~\cite{llmstudy}. Furthermore, determining the length of the output sequence in advance is usually impracticable, as the auto-regressive generation process continues until either the \textit{EOS} token is generated or the token limit is reached.

\begin{table}
\caption{The memory capacity and inference time across different GPUs. The time denotes the iteration time used to go through all layers, which is profiled from serving a batch of request under an OPT-2.7b model. The number of requests served in the prefill and decode phases are 3 and 25}
\centering
\small
\begin{tabular}{|c|c|c|c|}
\hline
Device & Memory & Time (Prefill) & Time (Decode) \\
\hline\hline
A100 & 80 GB  & 0.06s & 0.0097s \\
\hline
3090 & 24 GB  & 0.147s & 0.0143s \\
\hline
P100 & 12 GB  & 1.47s & 0.077s  \\
\hline
\end{tabular}
\label{table:memory_compute_gap}
\end{table}

\subsection{GPU Heterogeneity}
With the rapid evolution of GPU technologies, the computation power among GPUs are increasingly significant. As recent studies have extensively demonstrated, the computation power across GPUs gap yields significantly diverse acceleration effect for deep learning workloads~\cite{allox, gavel, heet, sia, gandivafair}. When it comes to LLM serving, the performance gap among GPUs becomes even more pronounced as the scale of model parameters, the volume of service requests, and the associated sequence length increase significantly. To illustrate this, we provide the memory and execution time information in Table~\ref{table:memory_compute_gap}, in which A100 offers  $1.47\times$ and $7.93\times$ higher inference capability than that under 3090 and P100 in decode phase while $2.45\times$ and $24.5\times$ in prefill phase.

Sufficient GPU memory is required to hold the model parameters, intermediate results, and KV caches during inference process.
However, the memory capacity gap among GPUs in clusters are also remarkable. As listed in Table.~\ref{table:memory_compute_gap}, A100 offers $3.33\times$ and $6.67\times$ greater capacity than 3090 and P100 respectively. This notable memory gap is of vital importance in LLM serving, since the limited memory capacity of low-end GPUs reduces the KV caches they can host and therefore bottlenecks their serving capability.

\subsection{Limitations of Existing Solutions}
\label{sec:limitations}
Given the inherent GPU heterogeneity in production clusters, state-of-the-art LLM serving systems have proposed various strategies to effectively utilize these heterogeneous GPU resources. Broadly, all these strategies are static and can be classified into two categories: one revolves around the splitting of heterogeneous GPU devices between prefill and decode phases, while the other centers on model parameter splitting based on the computational power gap across GPU devices. In this section, we explore the limitations of these static parallelization paradigms.

%
%

\begin{figure}[t!]
\begin{minipage}{0.49\linewidth}
\includegraphics[width=0.99\linewidth]{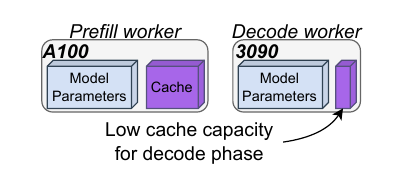}
\subcaption{{Phase-splitting}}
\end{minipage}
\begin{minipage}{0.49\linewidth}
\includegraphics[width=0.99\linewidth]{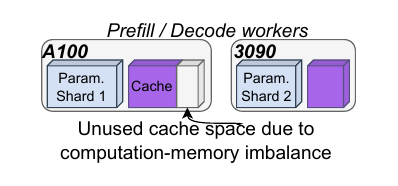}
\subcaption{{Parameter-splitting}}
\end{minipage}
\vspace{-1em}
\caption{Memory inefficiency in existing systems.}
\vspace{-1.5em}
\label{fig:memory_inefficiency}
\end{figure}

\textbf{Memory inefficiency}. Both phase- and model parameter-spliting schemes lead to GPU memory inefficiency in heterogeneous GPU clusters, severely impacting serving capability. First, separating the prefill and decode phases results in multiple copies of model parameters being replicated, like Splitwise~\cite{splitwise}, overwhelming accelerators with limited memory capacity. Consequently, there is a shortage of KV cache space on decode workers. As illustrated in Fig.~\ref{fig:memory_inefficiency}(a), when using an A100 and a 3090 GPU to host an FP16 7b model, only 10GB is available for KV caches on the decoding worker (i.e., the 3090 GPU), which adversely reduces decoding capacity. Furthermore, memory exhaustion in one phase inherently bottlenecks memory utilization gains in the other phase due to tight inter-phase coupling.

Second, the significant disparity between the memory and computational capabilities of heterogeneous GPU devices presents a fundamental challenge for Hexgen~\cite{hexgen}—a representative system that utilizes parameter-splitting. This imbalance prevents effective GPU memory utilization when employing parallelization schemes that focus solely on balancing the execution times of inference requests. For the example illustrated in Tab.~\ref{table:memory_compute_gap}, if we were to partition an FP16 OPT-2.7b model on an A100 and a 3090 GPU based on their respective computation capabilities during the decode phase under the fixed request arrivals, nearly 40\% of the parameters and KV caches would need to be located on the 3090 GPU, while 60\% would reside on the A100. In this scenario, the 3090 GPU would only have up to 18.6GB of available memory space for hosting the KV caches. Once the KV cache space in the 3090 GPU is fully exhausted, the serving system would be forced to stop hosting new requests, despite the fact that the A100 GPU still has a substantial 44.1GB of unused memory space as depicted in Fig.~\ref{fig:memory_inefficiency}(b). This issue becomes more critical under fluctuating request arrivals, as the performance gap between heterogeneous devices varies with batch sizes.

\begin{figure}[t]
\begin{minipage}{0.49\linewidth}
\includegraphics[width=0.99\linewidth]{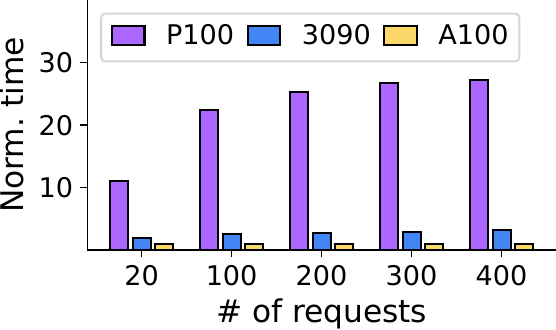}
\subcaption{MLP}
\end{minipage}
\begin{minipage}{0.49\linewidth}
\includegraphics[width=0.99\linewidth]{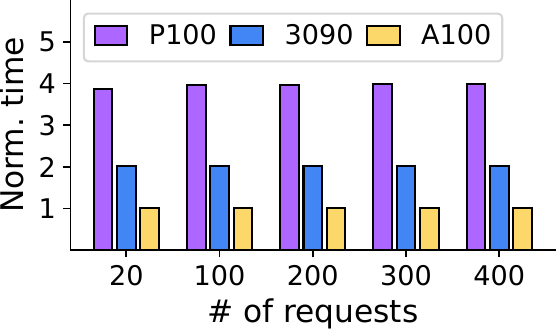}
\subcaption{Attention}
\end{minipage}
\caption{The execution time of the decoding MLP and Attention of one layer in Llama-70B across different GPUs. Sequence length of each request is set to 1000.}
\vspace{-1em}
\label{fig:execution_time_fixed_seqlen}
\end{figure}

\textbf{Computation inefficiency}. 
Splitwise struggles to utilize high-end GPUs during the decoding phase and low-end GPUs during the prefill phase, resulting in significant computational underutilization. This limitation is especially severe, as the P100 GPU lags behind the A100 by up to $40.4\times$ on average in MLP computation during decoding, as illustrated in Fig.~\ref{fig:execution_time_fixed_seqlen}(a). Moreover, this significant performance gap between the A100 and P100 undermines efforts to reduce latency through skewed computation partitioning in parameter-splitting systems, like Hexgen. Specifically, partitioning computations based on MLP computation gap results in only a small fraction of the workload being assigned to P100, offering minimal latency reduction. When multiple P100 GPUs are used alongside A100 GPUs, the latency improvement is easily offset by the increased communication overhead, which grows significantly with the number of GPUs. Furthermore, while the performance disparity between low-end and high-end GPUs narrows in the Attention module, as illustrated in Fig.~\ref{fig:execution_time_fixed_seqlen}(b), balancing the Attention computation between the A100 and P100 ultimately leads to prolonged MLP computation time on P100. This highlights the fact that achieving satisfactory inference performance is not feasible by partitioning the computations of all modules across GPUs in the same manner.

\textit{\textbf{Takeaway}}. Current heterogeneity-aware LLM serving systems face significant challenges in achieving low latency and high throughput, due to limitations such as restricted cache capacity and under-utilization of computational resources. Such inefficiencies stem from their coarse-grained and static parallelization approaches. Consequently, there is a critical need for a finer-grained and dynamic parallelization design to consistently meet the computational and memory requirements of inference service requests.

\subsection{Opportunities}
\label{sec:opportunities}
In light of the inefficiencies presented in existing approaches, we identify two opportunities that can be leveraged to ensure high-performance inference in heterogeneous GPU clusters:


\textbf{$\bm{O_1}$: Selectively distributing dense computation across GPUs benefits latency reduction.} While incorporating more GPU devices for LLM systems can accommodate additional KV caches—crucial for enhancing serving capacity—it can negatively impact inference latency. Specifically, low-end GPUs like the P100 significantly lag behind A100 in terms of performance when processing dense modules with high arithmetic intensity, such as MLP, proving negligible or even negative benefits in computation time reduction. However, it can introduce non-negligible communication overhead surges if these low-end GPUs are added, particularly in prefill phase that requires transferring large intermediate tensors across layers. Consequently, narrowing the scope of parallelization among heterogeneous GPUs—by selectively distributing these dense computations across a subset of GPU types—can help reduce inference latency and improve the resource utilization of high-end devices.

\textbf{$\bm{O_2}$: Parallelizing Attention computation among all devices benefits system throughput.} The Attention module is inherently parameter-free, meaning it operates solely on the input tensor $q$ and the KV caches without relying on any model parameters. This feature enables dynamic parallelization that is not restricted to specific GPU workers, allowing it to scale to adapt effectively to fluctuating request arrivals. As a result, the unused cache spaces created by asymmetric model sharding (as discussed in ~\cref{sec:limitations}) can be fully utilized after parallelizing the Attention computation to enhance throughput. Furthermore, the low computational intensity of the Attention module leads to comparable latency across all heterogeneous GPUs (as illustrated in Fig.~\ref{fig:execution_time_fixed_seqlen}(b)), facilitating effective load balancing strategies that help reduce computational overhead.

\section{Hetis System Overview}


\subsection{Key Design Ideas} 
Hetis is a new LLM serving system that leverages the above opportunities to minimize inference latency while achieving high throughput within heterogeneous GPU clusters. To achieve this, Hetis is built upon the following key ideas:

\textbf{$\bm{I_1}$: Optimized parallel configuration for dense operations.} Hetis identifies the optimal parallelization scheme among heterogeneous devices for compute-intensive computations, i.e., dense computations and prefill Attention, to optimize the inference performance. Specifically, a balance between the computation latency and communication overhead is achieved, where possibly only partial GPUs are included as a consequence.

\textbf{$\bm{I_2}$: Dynamic head-wise Attention parallelization.} Hetis enables flexible KV cache storage and computation distribution across all heterogeneous GPUs during Attention computation. Specifically, it introduces a dynamic, head-level Attention parallelization mechanism that efficiently utilizes cluster resources while adapting to fluctuating request arrivals, in a communication-efficient manner.

\textbf{$\bm{I_3}$: Global request scheduling via explicit latency quantification.} Hetis implements an online scheduling policy built on dynamic Attention parallelism to allocate the optimal number of heads to each GPU device for all inference requests. This policy opportunistically leverages low-end GPU devices to minimize latency gaps among Attention workers by explicitly modeling both computation and communication overhead of Attention operations.

\begin{figure}
\centering
\includegraphics[width=0.95\linewidth]{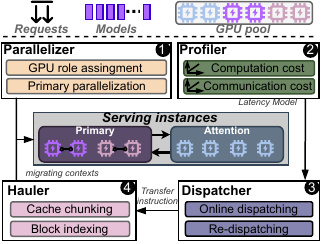}
\caption{The system architecture of Hetis.}
\vspace{-1em}
\label{fig:hetis}
\end{figure}

\subsection{System Architecture} 
As illustrated in Fig.~\ref{fig:hetis}, Hetis integrates four main modules to achieve its design goals: the Parallelizer, Profiler, Dispatcher, and Hauler. During system initialization, the Parallelizer~\ding{182} assigns distinct roles—namely, Primary and Attention workers—to heterogeneous GPUs to maximize resource efficiency within the cluster. Specifically, the Primary worker handles all operations during both the Prefill and Decode phases of inference, while Attention workers are dynamically pooled to opportunistically perform decoding Attention tasks. Simultaneously, the Parallelizer determines the optimal parallelization scheme among Primary workers by utilizing data-, pipeline-, and tensor-parallelism. In the meanwhile, the Profiler~\ding{183} conducts a lightweight yet effective profiling process to model computation and network cost for distributed computation of both MLP and Attention modules under various conditions. Leveraging this knowledge, Hetis is capable of making online decisions that dynamically distributes the Attention heads of requests through the Dispatcher~\ding{184}. The Dispatcher optimizes Attention time by scheduling computation and memory loads of new requests across all Primary and Attention workers in real-time.  It also mitigates potential service degradations by load balancing on-going requests at minimal cost when necessary. Furthermore, Hauler~\ding{185} implements efficient KV cache management at the head granularity, enabling zero-overhead migration of KV cache across GPUs.

\begin{figure}[!t]
\centering
\includegraphics[width=0.9\linewidth]{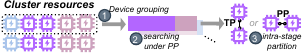}
\caption{The hierarchical searching process for configuring primary worker parallelism.}
\vspace{-1em}
\label{fig:hetis_primary_exploration}
\end{figure}

\section{Parallelization Scheme Design}
In this section, we delve into the design of Hetis' fine-grained and dynamic parallelization strategies. Specifically, Hetis introduces primary worker parallelism and dynamic Attention parallelism to collaboratively optimize dense and Attention computations.

\subsection{Primary Worker Parallelism}
To minimize the inference latency of dense computation, the Parallelizer identifies the optimal parallel configuration, denoted by $\sigma^*$, from a selected subset of devices drawn from a pool of available GPUs. This configuration is used to execute the model $M$ for a given inference request distribution $\mathcal{R}$, which encapsulates information about batch size and sequence length.  To be specific, the optimization problem can be formulated as:
\begin{equation}
\sigma^* = \arg\min_{\sigma}\ C(\sigma, M, \mathcal{R}).
\label{eq:primary_objective}
\end{equation}
Here, the mapping $\sigma$ assigns the entire model parameters to specific devices, supporting data-parallelism (DP), tensor-parallelism (TP), and pipeline-parallelism (PP) across heterogeneous GPUs. Furthermore, the optimal configuration $\sigma^*$ minimizes the total cost $C(\cdot)$, which is defined as the sum of the communication cost $C_{\text{comm}}(\cdot)$ and the computation cost $C_{\text{comp}}(\cdot)$~\cite{hexgen} for dense modules.

However, this optimization problem is NP-hard~\cite{hexgen}, and finding $\sigma^*$ is computationally expensive due to the vast search space. This challenge is further compounded by the fact that not all GPUs in the cluster need to participate in dense computations. To address this, the Parallelizer employs a hierarchical exploration process, enabling rapid searching while ensuring high resource efficiency. As illustrated in Fig.~\ref{fig:hetis_primary_exploration}, the process begins by grouping devices to form LLM serving instances, where GPUs of different types are evenly divided across all instances. Note that configurations without sufficient KV cache spaces to host the decoding process of $\mathcal{R}$ are filtered out. This approach significantly reduces the search space, as the low latency can be effectively sustained when applying load balance among data-parallel instances in clusters. 
Next, the Parallelizer optimizes model partition within each device group by leveraging PP, treating GPUs of the same type as a unified pipeline stage. To efficiently construct the mapping from layers to stages, it minimizes $C_p(\cdot)$—the maximum computation cost across all stages—under the assumption of perfect latency scaling~\cite{heet}, without accounting for communication overhead. Leveraging the fact that low-end GPUs are more resource-efficient for decoding Attention computation, the Parallelizer employs a searching heuristic to exclude GPUs that will be used solely for Attention computation, without compromising the computational efficiency of dense modules. Specifically, GPUs $\kappa$ are removed one by one, from the lowest- to highest-end type (in terms of computational power) within each device group, if the following criterion holds:
\begin{equation*}
C_p(\sigma - \kappa, M, \mathcal{R}) \big/ C_p(\sigma, M, \mathcal{R}) \leq 1 + \Delta.
\label{eq:primary_objective}
\end{equation*}
Here, $\Delta$ is set to 0.05 by default. The rationale behind this criterion is that low-end GPUs, which contribute minimally to reducing the cost of dense computations, will be excluded from primary worker parallelism. Following this, the Parallelizer explores various combinations of TP and PP within each unified  pipeline stage in parallel to further reduce latency. This is achieved by adopting a modeling approach similar to Hexgen~\cite{hexgen} to quantify dense computation, specifically $C_{\text{comm}}(\cdot)$ for communication and $C_{\text{comp}}(\cdot)$ for computation, thereby evaluating each candidate configuration. The configuration that yields the lowest cost during this search process is selected as $\sigma^*$.
In each serving instance, the selected GPUs are designated as \textit{Primary workers}, while those that are not selected are referred to as \textit{Attention workers}. These Attention Workers are responsible for performing Attention computation and collectively serve as a shared pool, multiplexed by all Primary Workers simultaneously, for executing Attention operations and storing KV caches.


\begin{figure}[t]
\begin{minipage}{0.49\linewidth}
\includegraphics[width=0.99\linewidth]{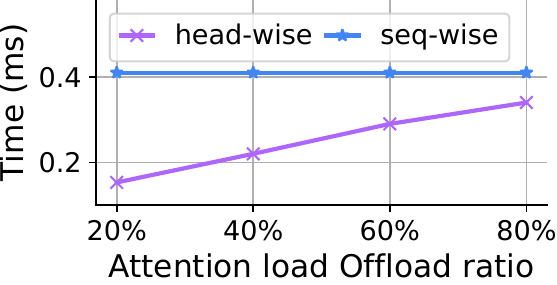}
\subcaption{Overhead under various loads}
\end{minipage}
\begin{minipage}{0.49\linewidth}
\includegraphics[width=0.99\linewidth]{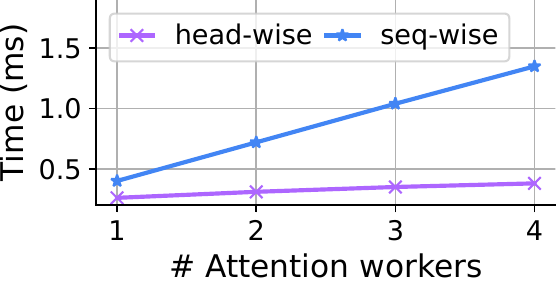}
\subcaption{Overhead across workers}
\end{minipage}
\caption{The advantage of head-wise splitting in terms of communication overhead. Experiments are conducted on Llama-70B models over 100Gbps network.}
\vspace{-1em}
\label{fig:head_wise_advantage}
\end{figure}

\subsection{Dynamic Attention Parallelism} 
\label{section:dynamic_attention}

Although not involved in dense computation, Attention workers are well-suited for handling partial Attention computation. By leveraging this capability, Hetis opportunistically utilizes low-end GPUs at the module level, maximizing resource efficiency. Furthermore, to address highly fluctuating inference loads, Hetis implements a dynamic Attention parallelization mechanism that enables flexible offloading of Attention computations between Primary and Attention workers. This adaptability ensures efficient handling of dynamic request arrivals and varying sequence lengths, ultimately reducing latency and improving throughput.

While there is potential to fully utilize heterogeneous resources, partitioning Attention computation across GPUs can introduce non-negligible communication overhead, significantly impacting inference performance. In such cases, the inter-host communication overhead—primarily stemming from the scattering and gathering of intermediate representations—can outweigh the benefits of leveraging low-end GPUs. Consequently, determining the dimension along which to split the hidden states should be partitioned, which have the shape [\texttt{batch\_size}, \texttt{sequence\_length}, \texttt{hidden\_dimension}] during distributed inference, becomes a critical consideration.

\begin{figure}[!t]
\centering
\includegraphics[width=0.95\linewidth]{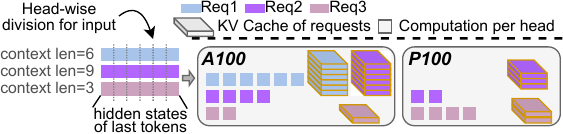}
\caption{Illustration of dynamic Attention parallelism: each request consists of six Attention heads and the context lengths of three requests are diverse.}
\label{fig:attention}
\vspace{-0.5em}
\end{figure}

First, partitioning Attention along the \texttt{batch\_size} dimension is too coarse-grained to provide precise control, as it distributes computation load and memory usage at the level of entire LLM inference requests. Consequently, this approach is ineffective in managing the unpredictable context length among requests, leading to frequent full-scale migrations of KV caches across GPU devices and resulting in significant GPU memory fragmentation. 
Second, dividing computation along the \texttt{sequence\_length} dimension requires transmitting the entire \textit{q} vectors to all decode workers that host partial caches of the request, and gathering the corresponding Attention values to continue subsequent computation. If a request spans multiple GPUs, its \textit{q} vector of the last token must be replicated and transferred multiple times. In contrast, partitioning the workload along the head dimension involves transferring only partial heads across GPU devices, which reduces the volume of communication. As illustrated in Fig.~\ref{fig:head_wise_advantage}(a), given one Attention worker, using head-wise partitioning can reduce the overhead by nearly $2.68\times$ if only $20\%$ loads are offloaded. 
Moreover, when more Attention workers are included to amortize the computation overhead, the head-wise splitting also demonstrates significant advantage. In Fig.~\ref{fig:head_wise_advantage}(b) where Attention loads of each request are evenly distributed to four workers, up to $3.55\times$ latency reduction is presented, as the head-wise communication pattern is more adept at avoiding network contention.

Overall, the head-wise Attention parallelism is formulated as:
\begin{subequations}
\begin{align}
&\mbox{Attention}_j = \mathsf{Concat}\big(\texttt{result}_{1,j}, \cdots, \texttt{result}_{n,j}\big), \\
&\mbox{and} \ \texttt{result}_{i,j} = \mathsf{softmax}\Big(q_{h^i_j(t)}\cdot K^T_{h^i_j(t)}\big/\sqrt{d}\Big)\cdot V_{h^i_j(t)}.
\end{align}
\label{eq:attention}
\end{subequations}
Here, $h^i_j(t)$ denotes the number of attention query heads, which represent the basic unit of the hidden dimension~\cite{attention}, allocated to request $j$ on GPU device $i$ at time $t$.
Fig.~\ref{fig:attention} provides an illustrative example of head-wise Attention parallelism, where different levels of Attention heads and KV caches are distributed among heterogeneous GPUs for various requests.

\section{Online Optimization}
In this section, we present Hetis' online optimization process, built on explicit modeling. This process dynamically schedules attention heads for inference requests across GPU devices within the Dispatcher, while opportunistically migrating KV caches between GPUs within the Hauler. 

\subsection{Modeling Dynamic Attention}
\label{sec:profiling}
\begin{figure}[t]
\begin{minipage}{0.315\linewidth}
\includegraphics[width=0.99\linewidth]{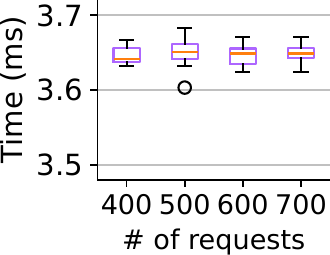}
\subcaption{Num. of requests}
\end{minipage}
\begin{minipage}{0.315\linewidth}
\includegraphics[width=0.99\linewidth]{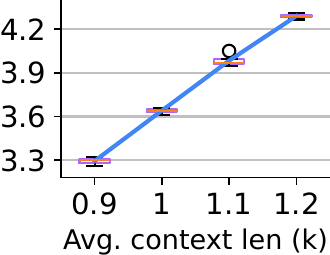}
\subcaption{context length}
\end{minipage}
\begin{minipage}{0.315\linewidth}
\includegraphics[width=0.99\linewidth]{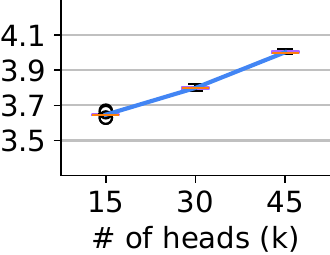}
\subcaption{Num. of heads}
\end{minipage}
\caption{Modeling of Attention computation time for OPT-30B. (a) Attention time remains independent of the number of requests, when the total number of heads and cache size  among requests are fixed. (b) Attention time grows linearly with the size of cache space. (c) Attention time grows linearly with the number of heads.}
\vspace{-0.5em}
\label{fig:execution_time_modeling}
\end{figure}

The dynamic Attention parallelism expands the search space by offering individualized parallelizations for requests, necessitating a new modeling framework for precise prediction of Attention computation time. 
Additionally, we need to consider the network overhead of distributed Attention computation, primarily attributed to the transfer overhead involved in generating the attention result.  To address these issues, Hetis adopts head-granularity modeling since the Attention time remains consistent regardless of the \texttt{batch\_size}, provided the number of query heads and cache space among requests are fixed (Fig.~\ref{fig:execution_time_modeling}(a)).

The Attention time in the decoding phase consists of several components: the vector-matrix multiplication between incoming $q$ vectors and the previously stashed $K$ and $V$ matrices, the overhead of transferring caches from GPU memory to GPU SRAM \cite{flashattention}, and the interference among Attention computations from various heads. 
As shown in Fig.~\ref{fig:execution_time_modeling}(b), the Attention computation time increases linearly with the average context lengths of query heads from different requests. This is primarily because only the query vector of the last token for each request needs to attend to all previous keys and values during the decoding phase, making the computation overhead proportional to the size of the KV cache involved. 
Moreover, the HBM-SRAM transmission overhead is also proportional to the quantity of caches in principle. Furthermore, we investigate how the number of heads affects the Attention time. In Fig.~\ref{fig:execution_time_modeling}(c), the Attention time rises with the number of query heads given a fixed amount of cache space. This increase is due to heightened contention for hardware resources as more heads are involved. Consequently, the Profiler models the Attention computation time $\tau_i(t)$ on GPU device $i$ as a linear function of the number of query heads $h_i(t)$ among requests and the amount of associated caches  $g_i(t)$:
 \vspace{-.4em}
 \begin{equation}
 \tau_i(t) = a_i \cdot h_i(t) + b_i \cdot g_i(t) + c_i.
\label{eq:profiling1}
 \vspace{-.4em}
 \end{equation}
In this expression, $a_i$, $b_i$, and $c_i$ represent the respective parameters associated with device $i$, and $g_i(t)$ denotes the size of the overall cache spaces across heads from all requests located on GPU $i$.

We further analyze the transmission overhead between the Primary and Attention workers for head-wise chunked Q, K, V, and output vectors generated during Attention computation, leveraging the well-established linear Alpha-Beta model~\cite{alphabeta}. This model is particularly effective in capturing point-to-point transmission overhead. In the decode phase, only the last token of each request is used as input for Attention computation. Hence, the Profiler models the transfer overhead $\rho_i(t)$ between the Primary worker and any associated Attention worker $i$ using a linear function:
\begin{equation}
 \rho_i(t) = \gamma_i \cdot d_i(t) + \beta_i,
\label{eq:profiling2}
\end{equation}
where $\gamma_i$ and $\beta_i$ are the corresponding parameters determined through the offline profiling run, while $d_i(t)$ denote the volume of data required to transferred, including the querys, keys, values, and Attention results. Here, $d_i(t)=\big(2+\frac{2}{r}\big)\cdot h_i(t)$, where $r$ is the ratio between the number of query heads and the quantity of grouped key/value heads, allowing for seamless support of both GQA~\cite{llama} and MHA~\cite{opt} architectures.

\subsection{Dynamic Head-wise Dispatching}
\label{sec:scheduling}


\subsubsection{Head-wise division model}
We first define the dynamic parallelization problem within Hetis' Dispatcher module as follows: In each inference instance, every pipeline stage $k$ is composed of $\mathsf{N_k}$ primary and Attention workers. At any given time $t$, Hetis serves a set of requests, denoted by $\mathbb{S}(t)$. The context length of each request $j$ is characterized by $l_j(t)$ for all $j \in \mathbb{S}(t)$. The Dispatcher module in Hetis determines the parallelization variables $\big\{ x^{j}_1(t), x^{j}_2(t),\cdots, x^{j}_{N_k}(t) \big\}$ for all requests, where $x^j_i(t)$ denotes the number of query heads of request $j$ allocated to the $i$-th GPU in the $k$-th stage. Note that $x_i^j(t) / r \in\mathbb{N}$ must always hold, as fractional numbers of head groups have no practical meaning. To ensure that Dynamic Attention parallelism does not affect the Attention workflow, we enforce the head-level integrity constraint as:
\begin{equation}
\small
\sum_{i=1}^{\mathsf{N_k}} x_i^j(t) \in \{0, \mathsf{H} \}, \ \forall j\in \mathbb{S}(t), \ \ \forall t.
\label{eq:con1}
\end{equation}
Here, $\mathsf{H}$ is the specified number of query heads for each request in this model. Furthermore, it is essential to ensure that hosting partial keys and values of new requests does not exceed the available cache capacity $\mathsf{M_i}(t)$ of all GPUs within this stage:
\begin{equation}
\small
\sum_{j\in\mathbb{S}(t)} x_i^j(t) \cdot l_j(t) \leq \frac{r\cdot\mathsf{M_i}(t)}{2},\ \ \forall 1\leq i \leq\mathsf{N_k}, \ \ \forall t.
\label{eq:con2}
\end{equation}

The objective of dispatching is to minimize Attention time, including network transfer overhead, for all requests by dynamically managing head-wise parallelization of Attention computation. This can be reduced to a conventional job-shop scheduling problem~\cite{garey1976complexity}, which is known to be NP-Hard.


\subsubsection{Efficient dispatching algorithm}
To facilitate rapid dispatching and minimize cache transfer across GPUs, the Dispatcher does not re-parallelize requests that have already been dispatched when distributing heads among GPU devices for newly incoming requests. This approach reduces the time spent navigating a large optimization space and prevents inference blocking due to extensive cache transfers, thereby avoiding unnecessary delays. 

 Since the total Attention time is determined by the maximum Attention time on GPUs during the post-Attention aggregation among workers within each serving instance, the optimization problem can be formulated as follows:
\begin{subequations}
\small
\begin{align}
\min \max_{i} & \ f_i\big(\vec{x_i}(t)\big),   \\       
\mbox{s.t.,}    & \ g_i(t) + \sum_{j=1}^{\mathsf{J}(t)} x^j_i(t)\cdot l_j(t) \leq \frac{r\cdot\mathsf{M_i}(t)}{2}, \ \forall i, t, \\
& \ \sum_{i=1}^{\mathsf{N_k}} x_i^j(t) = \mathsf{H},\ \forall j\in\mathsf{J}(t).
\end{align}
\label{eq:optimization1}
\end{subequations}
Here, $\mathsf{J}(t)$ represents the set of newly incoming requests at time $t$, $\vec{x_i}(t)=\Big\{ x^1_i(t),  x^2_i(t), \cdots, x^{|\mathsf{J}(t)|}_i(t)\Big\}$ is a per-device vector that includes the dispatching variables for all incoming requests on device $i$ at time $t$, and $f_i(\cdot)$ represents the execution time required for computing Attention. Specifically, for the Primary workers, we have $f_i\big(\vec{x}_i(t)\big) = a_i \cdot \big( h_i(t) + \sum_{j=1}^{\mathsf{J}(t)} x^j_i(t)\big) + b_i \cdot \big(g_i(t)+ \frac{2}{r}\cdot\sum_{j=1}^{\mathsf{J}(t)}  l_j(t)x^j_i(t) \big) + c_i$. For Attention workers involving network transfer with primary workers, we have $f_i\big(\vec{x}_i(t)\big)= \big( a_i + (2+\frac{2}{r})\cdot\gamma_i\big) \cdot \big( h_i(t) + \sum_{j=1}^{\mathsf{J}(t)} x^j_i(t)\big) + b_i \cdot \big(g_i(t)+ \frac{2}{r}\cdot\sum_{j=1}^{\mathsf{J}(t)} l_j(t)x^j_i(t) \big) + c_i + \beta_i$.

Given the linear nature of attention mechanisms and the point-to-point communication overhead, the optimization problem can be efficiently reformulated as a linear programming problem, allowing for a polynomial-time solution~\cite{LP}. After solving the problem at time $t$, the values of $h_i(t)$ and $g_i(t)$ are updated based on the dispatching solution $\vec{x_i}(t)$:
\begin{equation}
\small
h_i(t+1)  = h_i(t) + \sum_{j=1}^{\mathsf{J}(t)} x^j_i(t); g_i(t\text{+}1)  = g_i(t) + \frac{2}{r}\cdot\sum_{j=1}^{\mathsf{J}(t)} x^j_i(t)\cdot l_j(t).
\end{equation}

\subsection{Re-dispatching Design}
The dispatching strategy described above may struggle to maintain high efficiency during the decoding of unpredictable long requests. To address this, the Dispatcher integrates a re-dispatching scheme into Hetis to manage already dispatched requests, ensuring token generation latency is safeguarded while enhancing serving capacity consistently.

\subsubsection{Balance computation time}
Since the execution time of the Attention module in Hetis is dictated by the maximum computation time among GPUs, requests with unpredictably long contexts can significantly extend the Attention time. Specifically, the consistent increase in context length for certain long requests can create highly imbalanced Attention loads, undermining efforts to minimize Attention time by dispatching heads for newly arrived requests. Consequently, the re-dispatching scheme aims to balance computation loads and opportunistically reduce the execution time of the Attention module for all existing requests, represented by $\mathbb{S}(t)$. To achieve this, the Dispatcher of Hetis first computes the ideal Attention time $f^*$ under the optimal head-wise dispatching scheme within the serving instance  through solving the following optimization problem:
\begin{equation*}
\small
f^* = \min \max_i f_i\big(\vec{x_i}(t)\big); \ \mbox{s.t.},\ \sum_{i=1}^{\mathsf{N_k}}\sum_{j=1}^{\mathbb{S}(t)}x^j_i(t)l_j(t)\leq\sum_{i=1}^{\mathsf{N_k}} \frac{r\cdot\mathsf{M_i}}{2}.
\label{eq:ideal_time}
\end{equation*}
If the deviation between ideal Attention time and the current Attention time exceeds a threshold $\Theta$, set at 50\% by default, Hetis selectively re-dispatches one request that offers the greatest potential for reducing Attention time, aiming to balance execution time reduction with decoding interruptions. Specifically, Hetis first identifies the device $i$ with the longest Attention computation time, which serves as the bottleneck for the entire Attention module. Next, it locates the request $j$ that contributes most significantly to the Attention load at device $i$.
Hetis then re-dispatches this request by dispatching its attention heads among heterogeneous GPUs according to the formulation in Eq.~\eqref{eq:optimization1}.

\subsubsection{Balance KV cache}
The exhaustion of cache space on GPUs hinders the token generation process for requests that continuously require free memory to store the key and value vectors for newly generated tokens. Previous studies~\cite{pagedattention} have primarily relied on request preemption schemes, such as LIFO (last-in-first-out) and LRU (least-recently-used), both adapted from traditional cache management practices to determine the eviction order of requests. However, these approaches rest on a critical assumption: that GPUs have uniform memory space and that the memory consumption of requests is identical across devices. Consequently, applying such eviction strategies may be ineffective for releasing cache space on memory-exhausted GPUs in Hetis, as preempted requests may not actually consume cache space on the overloaded devices. To address this limitation, Hetis narrows the search space to include only those requests that consume memory resources on the target GPU devices. For instance, Hetis modifies the LIFO approach to evict the request with the latest arrival time on the memory-exhausted device. Moreover, instead of fully discarding or swapping out the KV caches, Hetis opportunistically utilizes the available GPU memory within a cluster. Specifically, Hetis first checks whether $\sum_i g_i(t) < \sum_i \frac{r\cdot\mathsf{M_i}}{2}$ holds, where a positive result indicates that re-dispatching the victim request among GPU devices, according to Eq.~\eqref{eq:optimization1}, can alleviate memory exhaustion. This re-dispatching process continues until there is no available memory in the cluster or the gap with the ideal computation time exceeds the specified threshold $\Theta$. 

Notably, the re-dispatching design leverages the overlap in head distribution between the old and new parallelization schemes to opportunistically reuse the cache. As a result, Hetis minimizes cache migration overhead by implementing only partial cache transmission, facilitated by head-wise cache management introduced in~\cref{sec:implementation}.

\section{System Implementation}
\label{sec:implementation}
We have developed Hetis as an extension of the widely-adopted open-source vLLM framework~\cite{pagedattention}, which provides versatile inference functionality and high extensibility.

\noindent\textbf{KV cache management.} Similar to vLLM, we manage the memory usage of KV caches at the granularity of fixed-sized blocks. However, to support dynamic head-wise computation parallelization across GPUs, we implement a more fine-grained cache management scheme, which further splits cache blocks on the head dimension. To this end, we develop new CUDA kernels to fetch and store caches via the combination of \texttt{sequence id}, position within the sequence, and \texttt{head id}. Furthermore, since KV caches are managed at the head granularity—meaning each token per request has multiple caches on each GPU—we accelerate this compute-intensive block indexing process during the decode phase by leveraging multi-core parallelization on the CPU.

\noindent\textbf{Attention computation.} Since the Attention computation is orthogonal to the head dimension, we can parallelize the computation without needing to aggregate global \texttt{softmax} attributes. To this end, we reuse the PagedAttention kernel~\cite{pagedattention} to compute the partial Attention results based on the heads located at each device. 

\noindent\textbf{Communication optimization.} Dynamic Attention parallelism complicates the communication paradigm among workers. To address this, we implement it by additionally setting new communication groups within NCCL~\cite{nccl}, which allows for peer-to-peer (P2P) transmission between the Primary and Attention workers. 

\noindent\textbf{Live cache migration.} The live migration of KV cache can concurrently occur with collective communication, which introduces severe resource contention and prolongs the inference latency. To reduce this adverse impact on the inference process, the cache migration in the Hauler is operated on low-priority CUDA streams, protecting collective communication operations for ongoing inference computation from being blocked.

\noindent\textbf{Optimization problem solving.} 
We employ the \texttt{cvxpy} library~\cite{cvxpy} to determine the optimal solutions of the Linear Programming formulations within the Dispatcher. This is achieved by leveraging the capabilities of the \texttt{MOSEK} solver, which is available within the \texttt{cvxpy} library.

\vspace{-.3em}
\section{Evaluation}
\subsection{Evaluation Setup}
\label{sec:setup}

\noindent\textbf{Cluster configurations}. We utilized a local heterogeneous GPU cluster consisting of the following hardware components by default: a host with four A100-80GB GPUs, two hosts with two NVIDIA 3090 GPUs on each, and a host equipped with four P100 GPUs. All hosts are interconnected via a 100Gbps LAN, while GPUs within each host are connected through PCIe channels.


\noindent\textbf{Workloads}. To emulate real-world serving, we examined three typical LLM applications, i.e., chatbot, code completion, and summarization. Specifically, we use ShareGPT~\cite{sharegpt} (SG) for chatbot application, HumanEval~\cite{humaneval} (HE) for the code completion task, and LongBench~\cite{longbench} (LB) for long-article summarization. The input and output length of requests exhibits substantial diversity among them. Furthermore, our evaluation encompasses a range of models, including Llama-13B, OPT-30B, and Llama-70B, with the latter being a GQA model and the others being MHA models.

\begin{figure}
\centering
\includegraphics[width=0.99\linewidth]{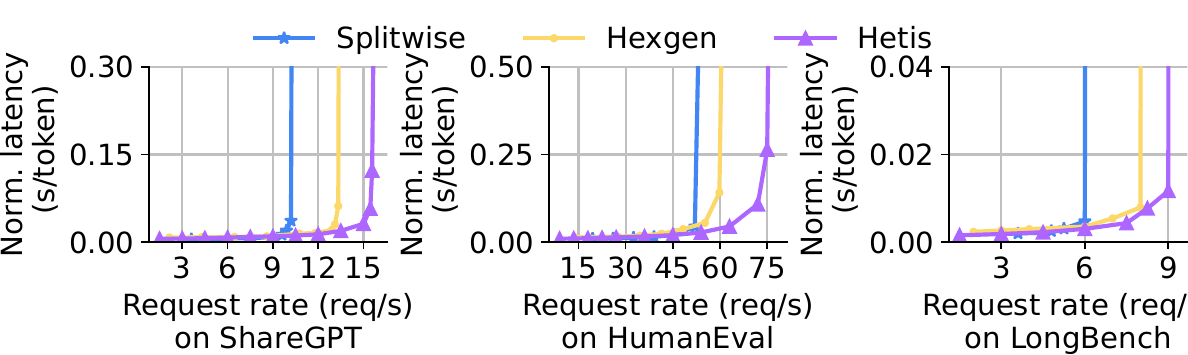}
\caption{{The normalized end-to-end latency across different datasets under Llama-13B.}}
\vspace{-0.5em}
\label{fig:13B_aggregated}
\end{figure}


\noindent\textbf{Baselines}. To demonstrate Hetis's ability to leverage heterogeneous GPUs under all types of existing parallelism, we compare it against the following heterogeneity-aware LLM serving engines:
\begin{itemize}
\item {\textbf{Splitwise}~\cite{splitwise}. It divides LLM requests into two distinct phases across heterogeneous GPU machines. Precisely, leveraging insights into the varying computational demands of the Decode and Prefill phases, it allocates prefill computations exclusively to high-end GPUs (such as A100) while processing the subsequent decode phase on low-end GPUs (like P100).}

\item {\textbf{Hexgen}~\cite{hexgen}. It employs static partitioning of computational loads for inference services on top of conventional model parallelism, utilizing both tensor and pipeline parallelisms. Hexgen allows for asymmetric parameter division to balance computation time across heterogeneous GPU devices. Unlike Splitwise, Hexgen places the prefill and decode phases on the same worker.}
\end{itemize}


\vspace{-0.5em}
\subsection{End-to-End Serving Performance}

In this section, we demonstrate the superior performance of Hetis through end-to-end experimental results, evaluating its effectiveness across three real-world workloads on heterogeneous GPUs.

\begin{figure}
\centering
\includegraphics[width=0.99\linewidth]{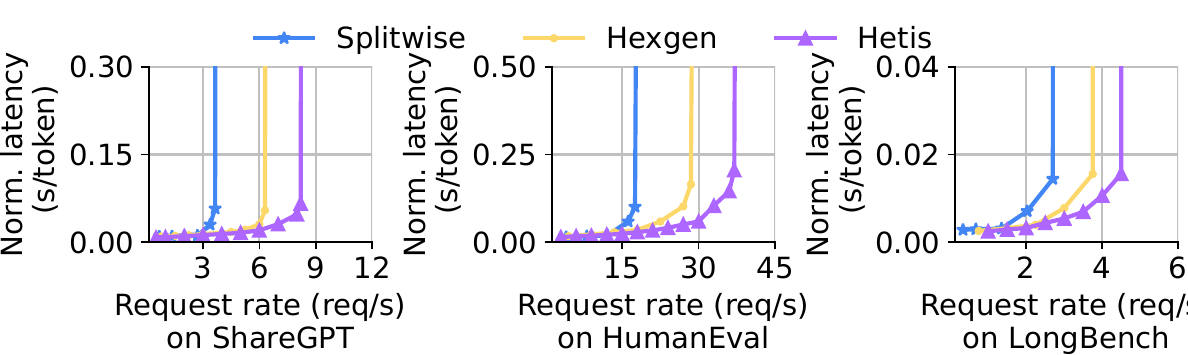}
\vspace{-1em}
\caption{{The normalized end-to-end latency across different datasets under OPT-30B.}}
\vspace{-1.5em}
\label{fig:30B_aggregated}
\end{figure}

\begin{figure}
\centering
\includegraphics[width=0.99\linewidth]{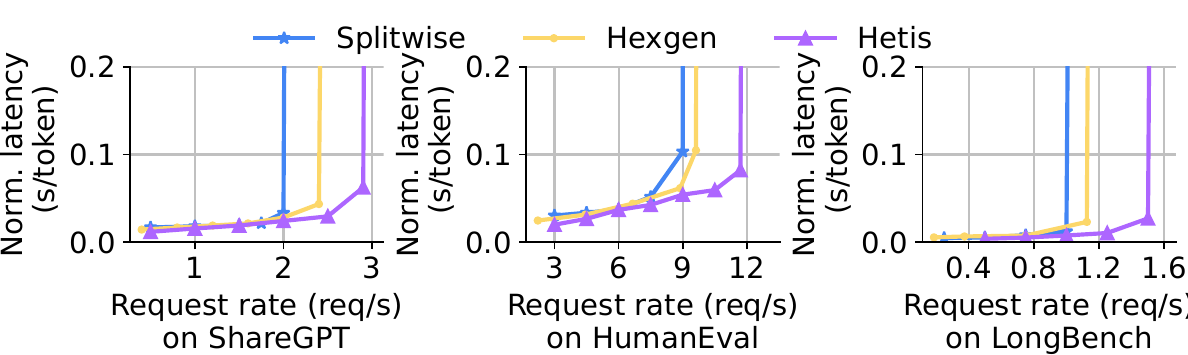}
\vspace{-1em}
\caption{{The normalized end-to-end latency across different datasets under Llama-70B.}}
\vspace{-1.5em}
\label{fig:70B_aggregated}
\end{figure}

We first evaluated model parallelism across baseline methods using the Llama-70B model. In Hetis, A100 and 3090 GPUs serve as Primary Workers, while P100s are dedicated to Attention Worker roles. Hexgen, in contrast, employs a four-stage pipeline parallelism, assigning homogeneous GPUs (e.g., A100s, 3090s, or P100s) to each stage and applying tensor parallelism within stages to optimize intra-stage computation.  Since Splitwise does not incorporate an explicit cluster-wide parallelism deployment strategy, we implemented it as a four-way TP prefill instance on A100 GPUs and a decoding instance consisting of two pipeline stages: one for two-way TP on 3090 GPUs and another for two-way TP on P100 GPUs. This design reflects the demand for high-end computational power across different phases of the workload. 

Under its novel parallelism design, Hetis consistently achieves a significantly higher service request rate  across various serving scenarios, as shown in Fig.~\ref{fig:13B_aggregated}, Fig.~\ref{fig:30B_aggregated}, and Fig.~\ref{fig:70B_aggregated}. Specifically, it demonstrates up to $2.25\times$ and $1.33\times$ higher throughput compared to Splitwise and Hexgen, respectively. This improvement can be attributed to Hetis's ability to prevent computational imbalances, optimize cache utilization, and minimize excessive cross-node communication—advantages that will be discussed in the sequel.

\begin{figure}
\centering
\includegraphics[width=0.99\linewidth]{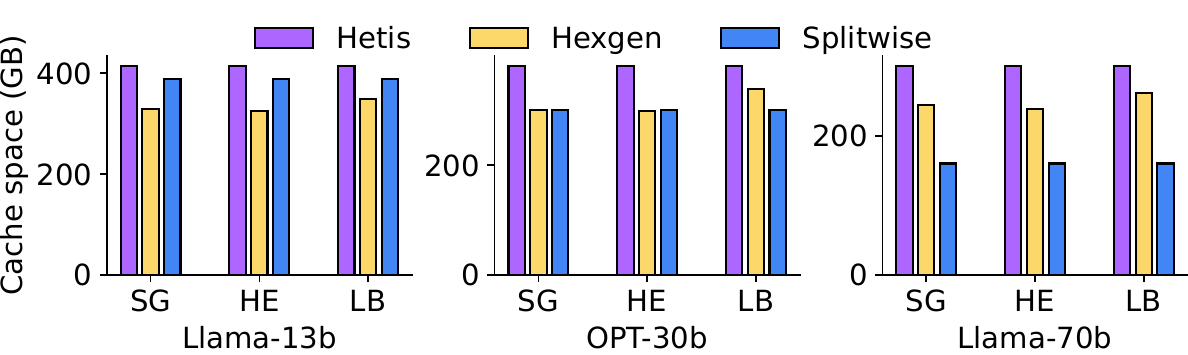}
\caption{The maximum available KV Cache space during inference across models and datasets.}
\vspace{-0.5em}
\label{fig:max_memory}
\end{figure}

We then analyzed the available cache spaces during inference across all baselines, which determine the maximum number of requests that can be hosted simultaneously in the cluster. As shown in Fig.~\ref{fig:max_memory}, Hetis consistently provides the largest amount of KV Cache blocks compared to the baselines, demonstrating up to $1.87\times$ improvement. In the case of Hexgen, its parallelization scheme varies based on request information, causing its available cache space to fluctuate across different datasets. However, due to the misalignment among heterogeneous GPUs in terms of computational and memory capabilities, Hexgen consistently underutilizes GPU memory, leading to significant waste even during peak times. In contrast, Splitwise dedicates excessive GPU memory to deploy multiple model parameters, which severely limits the remaining space available for hosting KV Caches. Even under the GQA model (e.g., Llama-70B), where KV Cache consumption is significantly reduced, the disaggregated design confines cache usage for each inference phase to their respective local devices. This restriction prevents Splitwise from achieving higher throughput.

\begin{figure}
\centering
\includegraphics[width=0.99\linewidth]{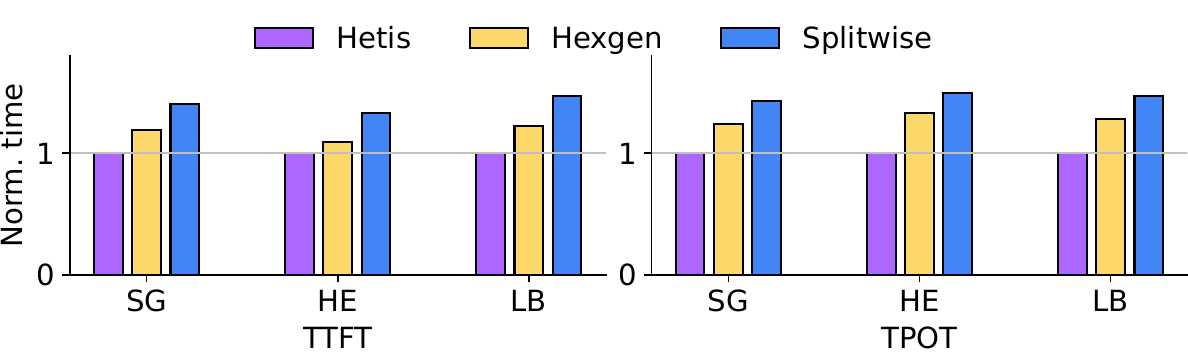}
\caption{The P95 TTFT and TPOT latency across datasets for Llama-70B models.}
\vspace{-0.5em}
\label{fig:p95latency}
\end{figure}

We further investigated the P95 TTFT (Time-to-First-Token) and TPOT (Time-Per-Output-Token) latency performance across baselines under the Llama-70B model, focusing on request rates of 1.5, 6, and 0.8 for ShareGPT, HumanEval, and LongBench, respectively—scenarios where none of the baselines are saturated by incoming requests. As shown in Fig.~\ref{fig:p95latency}, Hetis achieves up to $1.22\times$ and $1.47\times$ better P95 TTFT compared to Hexgen and Splitwise, respectively. This improvement is attributed to Hetis's ability to selectively involve partial GPUs for prefill computation, thereby reducing excessive communication overhead and avoiding latency prolongation caused by unbalanced parallelism.
In contrast, Splitwise suffers from full-scale transmission overhead due to the need to migrate KV Caches for each request and fails to opportunistically utilize the computational power of low-end GPUs, such as the 3090. As a result, it struggles to reduce prefill latency effectively. Hexgen, on the other hand, optimizes for end-to-end latency across all GPUs, but its inclusion of P100 GPUs in prefill computation introduces pipeline bubbles and increases communication overhead. This approach wastes the computational power of high-end GPUs and ultimately prolongs TTFT.

During the decoding phase, Hetis continues to maintain its advantages, with up to $1.39\times$ better TPOT over the baselines, as shown in Fig.~\ref{fig:p95latency}. Both Hexgen and Splitwise force low-end GPUs, such as P100s, to consistently handle dense computations during decoding, even under high inference loads. The significant computational gap in dense computations drags down the token generation process, outweighing the relatively smaller gap in Attention computations. Additionally, due to the prefill-decode disaggregation design, Splitwise is unable to leverage high-end GPUs effectively for decoding computations. By contrast, Hetis dynamically distributes computational loads across all types of GPUs while accounting for their respective resource capabilities. This design mitigates the adverse impact of relying on low-end GPUs for decoding, particularly their inefficiencies in dense computation, allowing Hetis to deliver consistently superior performance.


\subsection{Module-level Performance Evaluation}

To further understand the sources of improvement, we conducted an in-depth analysis of the execution time for the Attention and MLP modules—representative dense modules in Llama 70B—during the decoding phase. Experimental data are collected under the same request rates shown in Fig.~\ref{fig:p95latency}. Notably, due to the presence of pipeline parallelism across heterogeneous devices, the execution latency of specific modules within a token generation process may vary across different pipeline stages. To address this issue, we define the latency of a given module as the maximum execution time across stages, multiplied by the total number of stages. This metric reflects the module's contribution to token generation latency in the presence of pipeline bubbles~\cite{taming}.

\begin{figure}
\centering
\includegraphics[width=0.95\linewidth]{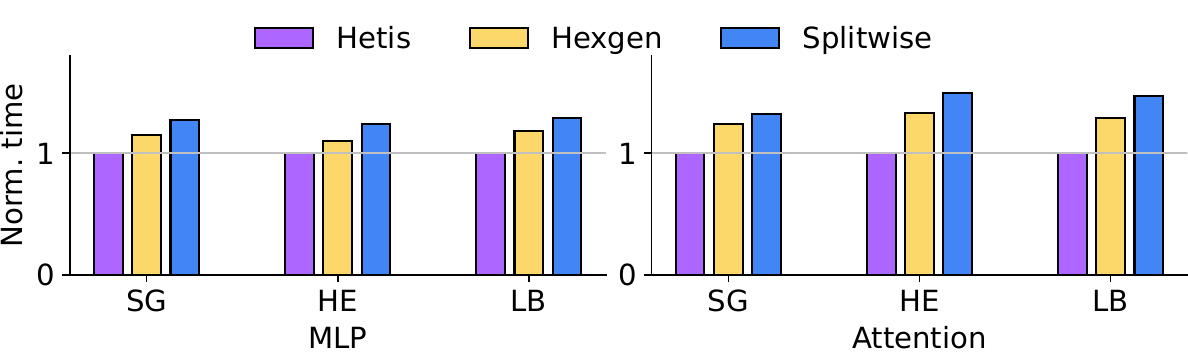}
\caption{The P95 execution latency on Attention and MLP modules during decoding phase across datasets for Llama-70B models.}
\vspace{-0.5em}
\label{fig:vllm_decoding_module_time}
\end{figure}

We begin by examining the MLP latency. As shown in the left part of Fig.~\ref{fig:vllm_decoding_module_time}, Hetis reduces MLP execution time by up to $1.29\times$. This improvement is attributed to the fine-grained parallel configuration explored by Hetis, which focuses on minimizing dense computation time rather than end-to-end latency. Notably, the experiments on the HumanEval dataset, which involve the largest number of decoding tokens during inference, demonstrate the most significant latency reduction.  Furthermore, thanks to the dynamic Attention parallelism design, we observe an even greater latency reduction for the decoding Attention operation—up to $1.49\times$, as shown in the right part of Fig.~\ref{fig:vllm_decoding_module_time}. Specifically, Hetis identifies the latency-optimal parallel configuration for Attention computation, consistently reducing Attention latency through both communication- and computation-efficient strategies.

\begin{figure}
\centering
\includegraphics[width=0.95\linewidth]{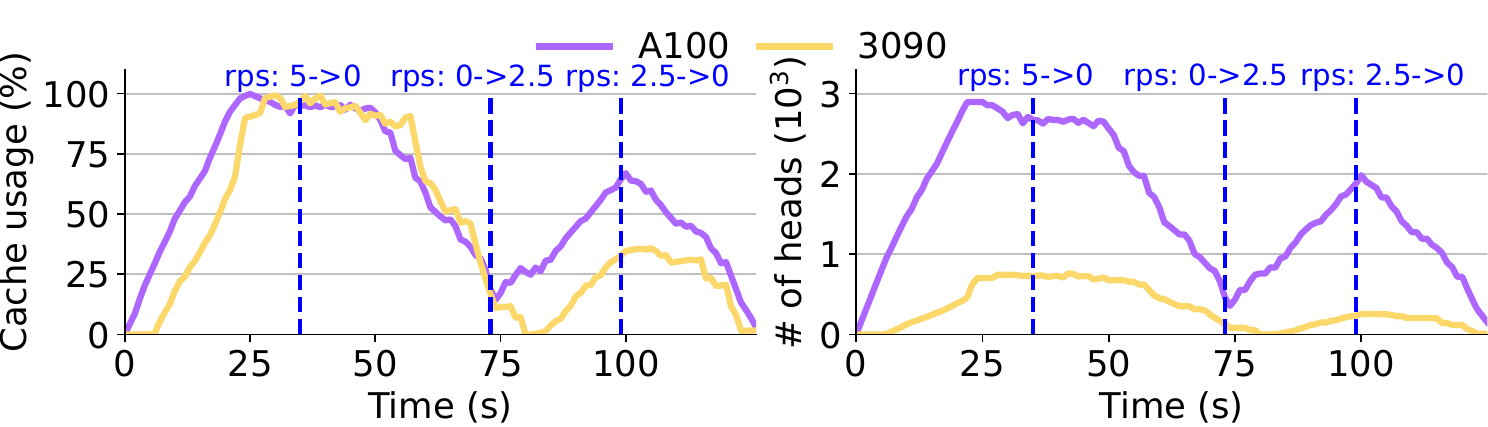}
\caption{Dynamic resource usage pattern in A100 and 3090 under time-varying request arrivals.}
\vspace{-0.5em}
\label{fig:varying_request_rate}
\end{figure}

To evaluate the efficiency of the dynamic Attention parallelism, we conducted an ablation study to analyze cache consumption and head distribution across A100 and 3090 GPUs over time, using requests from the ShareGPT dataset. For ease of analysis, we assigned the A100 as the Primary worker and two 3090 GPUs as Attention workers to serve the Llama-13B model. Fig.~\ref{fig:varying_request_rate} shows the fluctuations of both metrics under time-varying request rates. 
We observe that, due to the computational power gap and network partitioning, the A100 consistently handles more load than the 3090 GPUs (reflected by the number of heads), dynamically balancing the performance disparity between heterogeneous devices. Additionally, cache space on all devices is fully utilized during peak times, highlighting that Hetis avoids the memory inefficiency associated with static parallelism. 
Notably, at the start of serving, the 3090 GPUs begin processing Attention loads later than the A100. A similar trend is observed when the request rate increases from 0 to 2.5. This occurs because Hetis is aware of the additional overhead introduced by prematurely distributing loads across the network under light-loaded scenarios, thereby maintaining high efficiency.

\subsection{System Robustness and Overhead}
\label{microscopic-evaluation}

\noindent\textbf{Modeling accuracy}. We evaluated the profiling accuracy by comparing the estimations made by Hetis with the ground truth. Hetis conducts computation and network overhead modeling using the combination of eight $h_i(t)$ and eight $g_i(t)$. Thanks to the layer identity in LLM, Hetis only executes the Attention module once for each configuration, which typically takes no more than 100 milliseconds. Hetis consistently achieves high accuracy in predicting computation time across all GPU types, with accuracy levels reaching up to 93.8\%. Furthermore, we observe even higher accuracy in predicting transfer overhead, ranging from 92.4\% to 96.1\%.

\begin{figure}
\begin{minipage}{0.48\linewidth}
\includegraphics[width=0.99\linewidth]{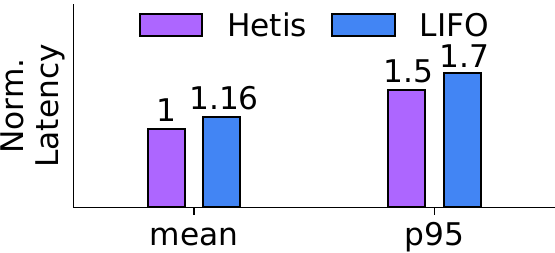}
\subcaption{Benefit of re-dispatching}
\end{minipage}
\begin{minipage}{0.48\linewidth}
\includegraphics[width=0.99\linewidth]{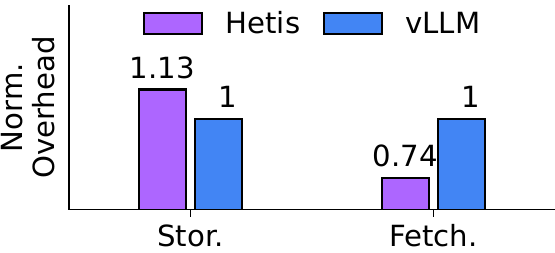}
\subcaption{Head management overhead}
\end{minipage}
\caption{{The benefit and management overhead  of dynamic head-wise dispatching scheme.}}
\vspace{-0.5em}
\label{fig:overhead2}
\end{figure}

\noindent\textbf{Benefit of re-dispatching}. We also examined the benefit of re-dispatching under Hetis on per-token latency, via comparing with the case where LIFO strategy is adopted. The experiment was conducted on ShareGPT dataset with request rate 5. Being aware of memory usage disparity, Hetis is capable to re-dispatch some requests on memory-exhausted GPU devices, thereby fully utilizing the memory in cluster. As shown in Fig.~\ref{fig:overhead2}(a), the mean output and P95 output latency is improved by $1.06\times$ and $1.14\times$ respectively.

\noindent\textbf{Searching overhead of primary worker parallelism.} In our heterogeneous GPU cluster, Hetis achieves a remarkably low searching overhead, generating the parallelization scheme in four seconds. Additionally, we evaluated its performance through a large-scale simulation featuring five GPU types with 32 GPUs each, where the search process completes in just 15 seconds. Since this process is only executed once prior to the system's initial deployment, its time cost is effectively negligible.

\begin{figure}[t]
\begin{minipage}{0.48\linewidth}
\includegraphics[width=0.99\linewidth]{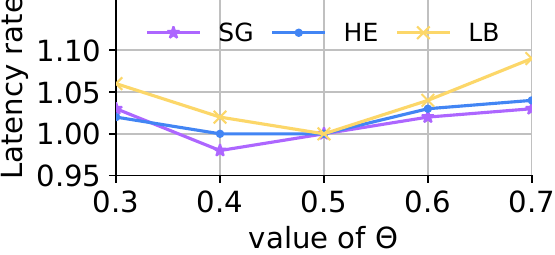}
\subcaption{Sensitivity to value of $\Theta$}
\end{minipage}
\begin{minipage}{0.48\linewidth}
\includegraphics[width=0.99\linewidth]{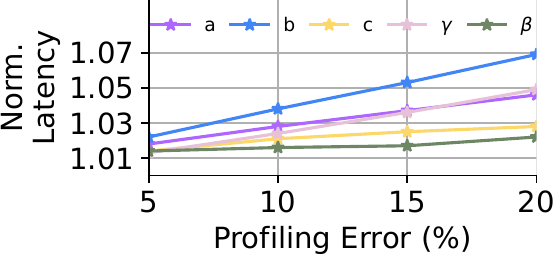}
\subcaption{Sensitivity to profiling error}
\end{minipage}
\caption{Robustness of Hetis system.}
\vspace{-0.5em}
\label{fig:sensitivity_analysis}
\end{figure}


\noindent\textbf{Head-wise cache management overhead}. We explored the overhead associated with changing the token-wise cache management in vLLM to a head-wise approach in Hetis by comparing cache space and fetching time. As shown in Figure~\ref{fig:overhead2}(b), the storage overhead in Hetis increases by 13\%, due to more storage operations being invoked with the same memory consumption for KV caches.
However, Hetis achieves a 26\% reduction in cache fetching time. This improvement is primarily due to the multi-core acceleration on cache block indexing process during the decoding phase.

\noindent\textbf{Re-dispatching factor control}. Selecting an overly large or small value of $\Theta$ adversely affects the inference process; a small value leads to excessive cache migration, while a large value results in imbalanced computation time. Fig.~\ref{fig:sensitivity_analysis}(a) shows that the default value under Hetis falls within an optimal range, achieving a balance across various benchmarks.

\noindent\textbf{Performance under inaccurate profiling.} Recognizing the potential for errors during the profiling phase, we also measured per-token latency of all requests under different levels of profiling error. Specifically, we varied the parameters $a_i$, $b_i$, $c_i$, $\gamma_i$, and $\beta_i$ introduced in Attention modeling (Section~\cref{sec:profiling}) by up to $\pm20\%$.
As shown in Fig.~\ref{fig:sensitivity_analysis}(b), even when the parameters deviated by as much as $\pm20\%$, the latency under Hetis witnessed a prolongation of only up to 6.9\%, highlighting the resilience to profiling errors.

\section{Related Works}

\noindent\textbf{LLM serving systems}. A substantial body of research has focused on developing LLM serving systems, each with diverse optimization targets such as latency, throughput, and cost~\cite{melange, hexgen, llmpq, spotserve, infinitellm, distserve, hu2024inference, orca, pagedattention, taming, muxserve}. Among these, Orca and vLLM serve as fundamental building blocks. Orca~\cite{orca} proposes continuous batching to improve GPU utilization, while vLLM introduces a novel memory-efficient PagedAttention kernel to mitigate memory fragmentation, thereby improving serving throughput from high memory utilization.

\noindent\textbf{CPU-GPU hybrid inference system}. Several studies have highlighted the impact of limited GPU memory capacity on LLM inference and proposed efficient CPU-GPU collaborative inference engines~\cite{flexgen, fastdecode}. By employing effective overlap designs that conceal the swapping overhead during extensive computations on GPUs, these approaches successfully enhance serving throughput. However, the costly CPU-GPU communication overhead is tailored for offline inference settings, which fundamentally undermines low-latency inference—contradicting our objectives.

\noindent\textbf{Heterogeneity-aware deep learning platforms}. Recognizing the significant acceleration potential across devices has drawn widespread attention to improving the efficiency of deep learning workloads in current datacenters. For instance, various schedulers~\cite{gandivafair, heet, sia} have been proposed to optimize GPU resource utilization and enhance training efficiency, while~\cite{hap, whale, metis} aim to accelerate training speeds in heterogeneous GPU clusters. However, there are currently no solutions specifically tailored to the dynamic characteristics of LLM serving, leaving considerable room for improvement.

\section{Conclusion}
This paper presents Hetis, a new heterogeneous-aware LLM serving system. Hetis features a fine-grained and dynamic parallelism architecture that enables precise control over heterogeneous resources at the module level. This design enhances memory, computation, and communication efficiency, dynamically adapting to the inference process and ensuring that resource demands of various modules of dynamic LLM requests are effectively met. Hetis further optimizes global resource distribution across all requests through a scalable head-level dispatching algorithm and a flexible re-dispatching strategy. Additionally, it seamlessly integrates advanced techniques such as GQA, leading to notable improvements in throughput and substantial reductions in inference latency.

\begin{acks}

We thank the anonymous reviewers for their helpful comments. This work is supported in part by the Science and Technology Development Fund of Macau (0071/2023/ITP2, 0024/2022/A1, and 0041/2025/RIA1), as well as the Multi-Year Research Grant of University of Macau (MYRG-GRG2024-00255-FST-UMDF and MYRG-GRG2023-00019-FST-UMDF).
\end{acks}

\bibliographystyle{ACM-Reference-Format}
\bibliography{paper}

\end{document}